\documentclass[aps,prl,twocolumn,showpacs,groupedaddress]{revtex4}

\usepackage{graphicx}

\begin{document}

\preprint{12}

\title{Novel Radiation-induced Magnetoresistance Oscillations in a Nondegenerate 2DES on Liquid Helium}

\author{Denis Konstantinov}
\email[E-mail: ]{konstantinov@riken.jp}
\author{Kimitoshi Kono}
\affiliation{Low Temperature Physics Laboratory, RIKEN, Hirosawa 2-1, Wako 351-0198, Japan}

\date{\today}

\begin{abstract}
We report the observation of novel magnetoresistance oscillations induced by the resonant inter-subband absorption in nondegenerate 2D electrons bound to the surface of liquid $^3$He. The oscillations are periodic in $B^{-1}$ and originate from the scattering-mediated transitions of the excited electrons into the Landau levels of the first subband. The structure of the oscillations is affected by the collision broadening of the Landau levels and by many-electron effects.    
\end{abstract}


\maketitle
\begin{figure}[t]
\centering
\includegraphics[width=8.0cm]{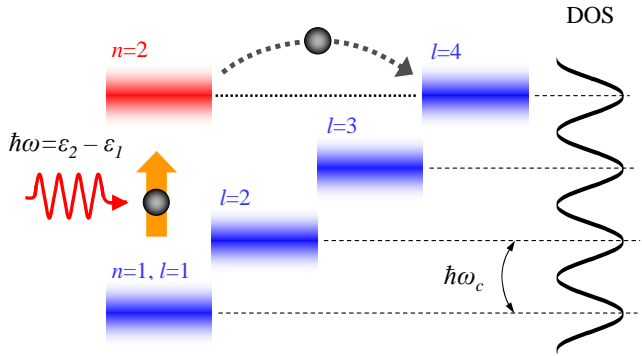}
\caption{\label{fig:2} (Color online) Schematic diagram of the single-electron energy levels. Levels labeled by $n=1$ (blue) and $n=2$ (red) correspond to the first and second energy subbands, respectively. Equidistant LLs of the first energy subband are labeled by index $l$. The broadening of levels is due to collisions of electrons with helium vapor atoms. The upward arrow indicates a resonant $n=1\rightarrow  2$ transition induced by radiation with angular frequency $\omega$. The dashed arrow indicates the inter-subband transition of an excited electron mediated by quasi-elastic scattering from a helium vapor atom. The solid line schematically shows the periodic modulation of the density of electron states due to the Landau quantization.} 
\end{figure}
\begin{figure}[b]
\centering
\includegraphics[width=8.5cm]{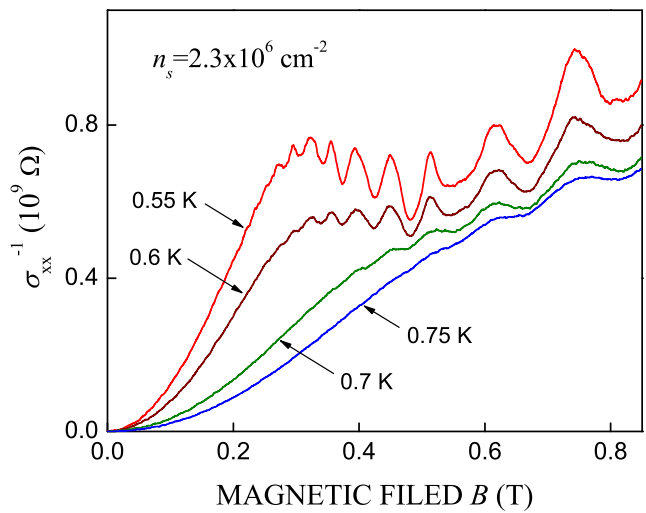}
\caption{\label{fig:3} (Color online) $\sigma_{xx}^{-1}$ vs $B$ for $n_s=2.3\times 10^6$~cm$^{-2}$ and for four different temperatures: $T=0.55$~K (red line), 0.6~K (brown line), 0.7~K (green line) and 0.75~K (blue line). All curves are for electrons irradiated with resonant 97~GHz microwaves at an input power of -5~dBm.}
\end{figure}
\begin{figure}[t]
\centering
\includegraphics[width=8.5cm]{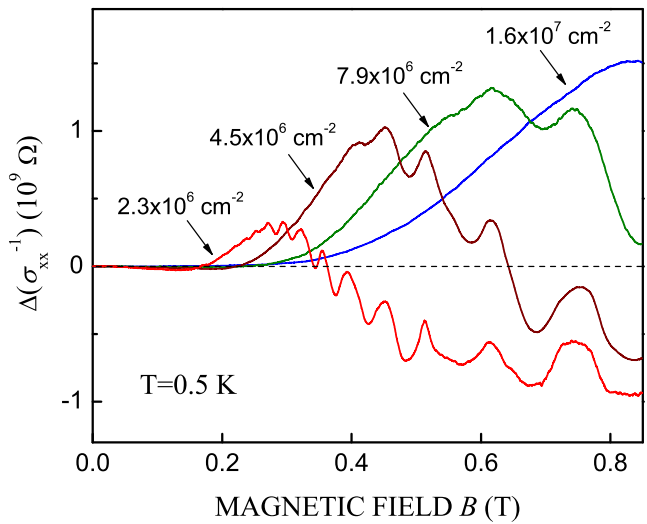}
\caption{\label{fig:4} (Color online) Change in the inverse conductivity $\Delta\sigma_{xx}^{-1}$ vs $B$ due to irradiation with 97~GHz microwaves at an input power of -5~dBm. The curves are taken at $T=0.5$~K and for four different electron surface densities: $n_s=2.3\times 10^6$~cm$^{-2}$ (red line), $4.5\times 10^6$~cm$^{-2}$ (brown line), $7.9\times 10^6$~cm$^{-2}$ (green line) and $1.6\times 10^7$~cm$^{-2}$ (blue line).}
\end{figure}

\indent The dynamical response of a two-dimensional electron system (2DES) is strongly affected by the magnetic field applied perpendicular to the 2D plane. In particular, the Landau quantization of the electron energy for the in-plane motion alters the transport properties of such systems and often results in oscillations of the electron magnetoresistivity. In degenerate 2DESs in semiconductors, the most well known example of such oscillations is the Shubnikov-de Haas oscillations arising from the sequential passing of Landau levels (LLs) through the Fermi level. Other examples include magnetointersubband oscillations in GaAs quantum wells~\cite{MamaniPRB2008} and microwave-induced resistance oscillations (MIRO) in GaAs/AlGaAs heterostructures~\cite{ZudovPRB2001, ManiNATURE2002}. A novel type of resistance oscillation induced by resonant inter-subband absorption in nondegenerate 2DES is the subject of this Letter.
\newline
\indent Low-density classical 2D electrons formed on the surface of liquid helium are the complement of quantum 2DESs in semiconductors~\cite{ColeRMP1974,AndreiBOOK}. The difference in energy between subbands is small and inter-subband transitions can be excited with resonant millimeter-wave radiation~\cite{GrimesPRL1974}. At high temperatures, the electron in-plane transport is determined by the short-range quasi-elastic scattering from helium vapor atoms, which is similar to the scattering due to disorder in semiconductors. At low temperatures (below 0.3~K for $^3$He), the scattering only occurs from surface capillary waves (ripplons). Unique correlation properties of 2DES are observed through the Wigner crystallization~\cite{CrandallPL1971}, quantum tunneling phenomena~\cite{SavillePRL1993} and inter-subband absorption~\cite{LambertPRL1980}.
\newline
\indent At high temperatures, the magnetotransport of 2DES on helium is well described by an independent-electron theory based on the self-consistent Born approximation (SCBA)~\cite{PetersPRB1994}. Below 1~K, the many-electron fluctuating electric field originating from the Coulomb interaction affects the electron scattering in both classically strong and quantizing magnetic fields. This was shown by Dykman \textit{et~al.}, who presented a theory for both ripplon~\cite{DykmanJETP1979} and vapor atom scattering~\cite{DykmanPRL1993}. Many-electron effects have also been observed in quantum cyclotron resonance~\cite{TeskePRL1999}.
\newline
\indent Interest in this system also stems from the proposals for quantum computing using electrons on helium ~\cite{PlatzmanSCIENCE1999,LyonPRA2006}. These proposals rely on the robust control of electronic quantum states using resonant microwaves. For such applications, microwave-induced inter-subband absorption has been recently studied for electrons on both liquid $^4$He and liquid $^3$He~\cite{CollinPRL2002}. It was shown that the continuous energy spectrum for the electron in-plane motion provides conditions for the strong heating of the electron system in such experiments~\cite{Ryvkine,KonstantinovPRL2007}. However, the effects of the in-plane quantization, which are particularly important for electronic qubits on helium~\cite{PlatzmanSCIENCE1999}, have never been investigated.
\newline 
\indent In this Letter, we present a study of inter-subband absorption in a 2DES on helium in the gas-atom scattering regime and in quantizing magnetic fields. Our main result is the observation of oscillations in the magnetotransport of electrons resonantly excited from the first subband to the second unoccupied subband. This phenomenon originates from the scattering-mediated transitions of the excited electrons into the LLs of the first subband. We show that inter-subband magnetotransport provides an effective method of studying many-electron kinetics in a 2DES with a quantized energy spectrum.  
\newline
\begin{figure}[b]
\centering
\includegraphics[width=8.5cm]{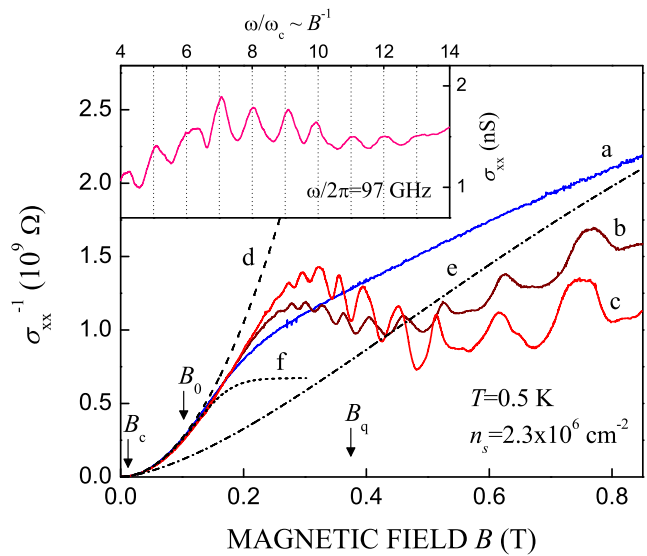}
\caption{\label{fig:1} (Color online) Inverse magnetoconductivity $\sigma_{xx}^{-1}$ vs $B$ for $n_s=2.3\times 10^6$~cm$^{-2}$, $T=0.5$~K without radiation (curve a) and for microwave powers of $-10$~dBm (curve b) and $-5$~dBm (curve c) at $\omega/2\pi=97$~GHz. Lines d (dashed), e (dash-dotted) and f (short-dashed) represent the Drude model, the single-electron theory and the many-electron theory for $\hbar\omega_c\ll k_B T$, respectively. The characteristic fields $B_{\bf c}$, $B_0$ and $B_{\bf q}$ (here $B_{\bf q}$ is such that $\hbar\omega_c=k_B T$) are indicated by arrows. Inset: magnetoconductivity $\sigma_{xx}$ versus $\omega/\omega_c$ at $\omega=\omega_{21}$ for $n_s=1.5\times 10^6$~cm$^{-2}$, $T=0.55$~K, $\omega/2\pi=97$~GHz and a microwave power of -5~dBm.} 
\end{figure}
\indent Electrons are bound on the free surface of liquid $^3$He placed approximately midway between two circular metal plates, which form a parallel-plate capacitor. A positive voltage is applied to the bottom plate to create an electric field $E_{\perp}$ perpendicular to the surface. In the asymmetric potential formed by the repulsive surface barrier, the attractive image force and the field $E_{\perp}$, the quantized energies $\epsilon_n$ ($n=1,2,...$) of the electron motion normal to the surface define the inter-subband transition frequencies $\omega_{n'n}=(\epsilon_{n'}-\epsilon_{n})/\hbar$. The inter-subband $n=1\rightarrow 2$ transition is excited using resonant microwave radiation of the angular frequency $\omega$ by tuning $\omega_{21}$ with the field $E_{\perp}$ through the linear Stark shift~\cite{GrimesPRL1974}.
\newline 
\indent The low-frequency (3~kHz) in-plane magnetoconductivity $\sigma_{xx}$~\cite{remark} is measured in magnetic fields of $B<1$~T applied perpendicular to the liquid surface. For these measurements, we employed the capacitive-coupling technique~\cite{SommerPRL1971} using a Corbino disk that constituted one of the circular plates. Qualitatively, the magnetotransport of 2DES can be described as a diffusion process with $\sigma_{xx}$ given by the Einstein relation~\cite{DykmanPRL1993}
\begin{equation}
\sigma_{xx}=\frac{n_s e^2}{k_B T}\frac{L^2}{\tau_{B}},
\label{eq:diff}
\end{equation}
\noindent where $n_s$ is the electron areal density, $L$ is the diffusion length and $\tau_B^{-1}$ is the scattering rate in the field. The Drude model assumes a continuous energy spectrum and a field-independent scattering rate $\tau_B^{-1}=\tau_0^{-1}$, where $\tau_0$ is the zero-field scattering time. However, in the field $B$ the electron energy spectrum is a discrete set of LLs, $\hbar\omega_c(l+1/2)$ (here $l=0,1,...$ and $\omega_c=eB/m$ is the cyclotron frequency), which are collision-broadened to a width of $\Delta_c=\hbar\tau_B^{-1}$. As a result, the scattering rate $\tau_B^{-1}$ is enhanced by a factor of about $\hbar\omega_c/\Delta_c$ as the electron states are concentrated from the range $\hbar\omega_c$ into $\Delta_c$. Therefore, we can determine $\tau_B^{-1}$ self-consistently from $\tau_B^{-1}\approx \tau_0^{-1}(\hbar\omega_c/\Delta_c)$. For a semi-elliptical density of states (DOS), the SCBA theory~\cite{AndoJPSJ1974} gives $\tau_B^{-1}=\tau_0^{-1}\sqrt{2\omega_c\tau_0/\pi}$. Correspondingly, in the single-electron approximation, $\sigma_{xx}$ deviates from the Drude model for fields $B\gtrsim B_{\bf c}$, where $B_{\bf c}$ is such that $\omega_c\tau_0=1$.
\newline
\indent The inverse $\sigma_{xx}$ versus $B$ measured for $n_s=2.3\times 10^6$~cm$^{-2}$ at $T=0.5$~K without radiation is shown in Fig.~\ref{fig:1} (curve \textbf{a}). Under such conditions, electrons occupy the first subband, and the population of higher subbands is negligible. For comparison, the Drude model and the single-electron theory are represented by lines \textbf{d} (dashed) and \textbf{e} (dash-dotted), respectively. The Drude conductivity can be obtained from Eq.~(\ref{eq:diff}) with $L^2=R_c^2/2$, where $R_c=\sqrt{2mk_B T}/eB$ is the classical cyclotron radius, and $\tau_B^{-1}=\tau_0^{-1}$. Line \textbf{e} is obtained from Ref.~10. Note that for $\hbar\omega_c\ll k_B T$, the single-electron conductivity can be obtained from Eq.~(\ref{eq:diff}) with $L^2=R_c^2/2$ and $\tau_B^{-1}\approx \tau_0^{-1}\sqrt{\omega_c\tau_0}$. In the quantum limit $\hbar\omega_c\gg k_B T$ we should use $L^2=l_B^2/2$, where $l_B=\sqrt{\hbar/eB}$ is the magnetic length.
\newline
\indent At low $B$, the observed $\sigma_{xx}^{-1}$ deviates significantly from that predicted by single-electron theory. This is due to the Coulomb interaction, which produces an in-plane electric field $\textbf{E}_f$ fluctuating in space and time and acting on each electron~\cite{DykmanJETP1979}. In the harmonic approximation, the probability distribution of $\textbf{E}_f$ is Gaussian with the r.m.s. field $\langle E_f \rangle\propto \sqrt{T_e}n_s^{3/4}$, where $T_e$ is the electron temperature. The field $\textbf{E}_f$ causes an ambiguity in the kinetic energy of a scattered electron during the collision event. This, in turn, affects the SCBA argument given earlier and leads to the reduction of the scattering rate~\cite{DykmanPRL1993}. The many-electron theory for $\hbar\omega_c\ll k_B T$ is represented in Fig.~\ref{fig:1} as line \textbf{f} (short-dashed). For $B<B_0$, where $B_0$ is such that $e\langle E_f \rangle R_c=\hbar\omega_c$, the variation of the electron energy across the diffusion length is larger than $\hbar\omega_c$, the Landau quantization is not significant and the Drude behavior is restored. At high $B$, where $\hbar\omega_c\gg k_B T$, the uncertainty $e\langle E_f \rangle l_B$ in the electron energy becomes much less than $\hbar\omega_c$, and $\sigma_{xx}^{-1}$ approaches the value predicted by single-electron theory. This behavior is in agreement with previous results~\cite{DykmanPRL1993}.
\newline
\indent However, the behavior of the conductivity under radiation markedly changes as shown in Fig.~\ref{fig:1}, where we plot $\sigma_{xx}^{-1}$ for input powers of $-10$~dBm (curve \textbf{b}) and $-5$~dBm (curve \textbf{c}). First, the conductivity deviates from its value without radiation. This will be discussed further in this Letter. An important novel feature is the appearance of oscillations periodic in $B^{-1}$. The origin of these oscillations can be understood by taking into account the inter-subband scattering of the excited electrons. A schematic diagram of the single-electron energy levels is shown in Fig.~\ref{fig:2}. For sufficiently large fields, electrons occupy only low-lying LLs of the first subband. When radiation is applied, the electrons are excited to the low-lying LLs of the second subband, and are subsequently scattered into the LLs of the first subband as a result of collisions with helium vapor atoms, as indicated by a dashed arrow in Fig.~\ref{fig:2}. Because these collisions are quasi-elastic, the electrons are scattered into states that have nearly the same energy as the initial states. Similar to the case of the intra-subband scattering discussed earlier, the rate of inter-subband scattering is enhanced because the electron states are concentrated into the broadened LLs. However, the important difference is that the number of the final states, the DOS of which is periodically modulated owing to the Landau quantization, depends on the relation between the inter-subband energy difference $\epsilon_2-\epsilon_1$ and the cyclotron energy $\hbar\omega_c$. As a result, we observe the variation of the scattering rate, and hence that of electron conductivity, with the ratio $\omega_{21}/\omega_c$.   
\newline
\indent The equilibrium distribution of electrons over LLs is attained through the electron-electron collisions. For $\hbar\omega_c\gtrsim k_B T$, the characteristic rate of this process was estimated to be $\omega_p^2/\omega_c$~\cite{DykmanPRL1993}, where $\omega_p=(2\pi e^2n_s^{3/2}/m)^{1/2}$ is the electron plasma frequency. For the moderate value of $B$ used in our experiment, this rate is much faster than the microwave-excitation rate. This allows us to introduce an effective electron temperature $T_e$, which is the same for all LLs. Note that the fast electron thermolization and the slow electron energy relaxation provide the conditions for the absorption-induced heating of the electron system~\cite{Ryvkine,KonstantinovPRL2007}.
\newline
\indent The proposed model gives an excellent account of the observed behavior. The inset of Fig.~\ref{fig:1} shows $\sigma_{xx}$ versus $\omega/\omega_c$ at $\omega=\omega_{21}$ for a 2DES irradiated with 97~GHz microwaves. The period of oscillations reflects the periodic modulation of the DOS due to the Landau quantization. In particular, the maximum DOS at $\omega/\omega_c=N$, where $N$ is a positive integer, results in the maximum of $\sigma_{xx}$ due to the enhancement of inter-subband scattering. Likewise, the minimum DOS at $\omega/\omega_c=N+1/2$ results in the minimum of $\sigma_{xx}$. Data obtained using microwaves with different $\omega$ showed similar scaling with $\omega_c$. The oscillations do not occur when $\omega$ is tuned away from $\omega_{21}$.
\newline
\indent The presented experimental method enables the direct observation of LL broadening due to collisions with helium vapor atoms. The number of vapor atoms strongly depends on $T$ and rapid increase in the scattering rate occurs as $T$ is increased. This leads to level broadening and to the smearing of the Landau spectrum. The effect of broadening can be clearly seen from the temperature dependence of oscillations. Figure~\ref{fig:3} shows $\sigma_{xx}^{-1}$ versus $B$ measured using electrons irradiated with resonant 97~GHz microwaves at $T=0.55$, 0.6, 0.7 and 0.75~K. As $T$ increases, the smearing of the Landau spectrum leads to the disappearance of oscillations starting from the low-field side where the interlevel separation is small. Above 0.8~K, oscillations are not observed in the range of applied fields because the level broadening restores the continuous spectrum for the electron motion along the plane.
\newline
\indent The effects of the many-electron fluctuating field $\textbf{E}_f$ are observed by varying the electron surface density $n_s$. Figure~\ref{fig:4} shows the radiation-induced change in the inverse conductivity $\Delta\sigma_{xx}^{-1}$ versus $B$ measured at $T=0.5$~K for four densities: $n_s=2.3\times 10^6$, $4.5\times 10^6$, $7.9\times 10^6$ and $1.6\times 10^7$~cm$^{-2}$. At low $B$, where electrons remain in the Drude regime, we observed a slightly negative $\Delta\sigma_{xx}^{-1}$. This is due to the heating of the 2DES by the absorbed microwaves and the increase in scattering as the higher excited subbands become thermally populated~\cite{KonstantinovPRL2007}. The enhancement of the scattering rate by about 10$\%$, determined using the Drude model, corresponds to $T_e\approx 3$~K. As $B$ increases, $\Delta\sigma_{xx}^{-1}$ becomes positive. This is explained by the increase in $\langle E_f \rangle$ due to electron heating and the corresponding decrease in the scattering rate due to many-electron effects~\cite{DykmanPRL1993}. The most notable feature of Fig.~\ref{fig:4} is the smearing of the oscillations with increasing $n_s$. This is due to the strong density dependence of $\langle E_f \rangle$. The $\textbf{E}_f$-induced ambiguity in the kinetic energy of a scattered electron increases with $n_s$. When this uncertainty in energy becomes comparable to $\hbar\omega_c$, an electron can scatter into the LLs of the first subband, regardless of the relation between $\omega_{21}$ and $\omega_c$. Therefore, at sufficiently large $n_s$ the oscillations disappear. We emphasize that this effect is different from the smearing of LLs due to collision broadening shown in Fig.~\ref{fig:3}. Actually, over a range of densities $n_s$, the many-electron effects can even lead to level narrowing, as was observed in cyclotron resonance~\cite{TeskePRL1999}. The smearing of the oscillations in Fig.~\ref{fig:4} results from the instantaneous tilting of the single-electron LLs in the in-plane fluctuating field $\textbf{E}_f$~\cite{DykmanPRL1993}. Interestingly, for low $n_s$ and at high fields, $\Delta\sigma_{xx}^{-1}$ becomes negative again. This may be due to the enhancement of the scattering rate due to the scattering-mediated transitions between the tilted LLs. A more detailed discussion of this result will be given elsewhere.  
\newline 
\indent Finally, we mention the relation between the described phenomenon and MIRO discovered in the quantum degenerate 2DES \cite{ZudovPRB2001,ManiNATURE2002}. The observation of zero-resistance states (ZRS) at the MIRO's minima triggered unprecedented theoretical interest in this effect~\cite{Fitzgerald2003}, but a complete explanation of MIRO and ZRS remains to be given. Remarkably, we observe similar ZRS induced by resonant inter-subband absorption in the nondegenerate 2DES on helium cooled to below 0.3~K. This new finding will be reported in a future publication. Here we emphasize that the study of novel resistance oscillations and ZRS in classical electrons on helium provides a new testing ground for microscopic theories of the effect of resonant radiation on disorder-scattered electrons.
\newline
\indent In conclusion, we have observed novel radiation-induced resistance oscillations in a nondegenerate classical 2DES on liquid helium. The oscillations originate from the scattering-mediated inter-subband transitions of the microwave-excited electrons and reflect the modulated structure of the density of electronic states due to the Landau quantization. The proposed model convincingly accounts for the dependence of these oscillations on various experimental parameters.\newline
\indent We acknowledge valuable discussions with M.~I. Dykman and Yu.~P. Monarkha. This work was supported in part by Grant-in-Aids for Scientific Research from MEXT.

%\bibliography{refdata}

\begin{references}
\bibitem{remark} Note that $\sigma_{xx}$ and the magnetoresistivity $\rho_{xx}$ are related to each other by $\sigma_{xx}=\rho_{xx}/(\rho_{xx}^2+\rho_{xy}^2)\approx \rho_{xx}/\rho_{xy}^2$, where the Hall resistance $\rho_{xy}=n_e e/B\gg \rho_{xx}$.
\end{references}
%

\end{document}